\documentclass[aps,twocolumn,amsmath,amssymb,floatfix,showkeys,
groupedaddress]{revtex4}

\usepackage{bm,amsmath,amssymb,amsfonts}

\usepackage[dvips]{graphicx}

\usepackage{color}
\definecolor{dred}{rgb}{0.50,0,0}
\begin{document}

\title{\textcolor{dred}{Engineering bands of extended electronic states 
in a class of topologically \\ disordered and quasiperiodic lattices}} 

\author{Biplab Pal}
\email{biplabpal@klyuniv.ac.in}

\author{Arunava Chakrabarti}
\email{arunava\_chakrabarti@yahoo.co.in}
\thanks{Tel.: +91 33 25820184, Fax: +91 33 25828282}

\affiliation{Department of Physics, University of Kalyani, Kalyani,
West Bengal-741235, India}

\begin{abstract}
We show that a discrete tight-binding 
model representing either a random or a quasiperiodic array of bonds, 
can have the entire energy spectrum or a substantial part of it absolutely 
continuous, populated by extended eigenfunctions only, when atomic sites are 
coupled to the lattice locally, or non-locally from one side. The event 
can be fine-tuned by controlling only the host-adatom coupling in one case, 
while in two other cases cited here an additional external magnetic field is 
necessary. The delocalization of electronic states for the group of systems presented 
here is sensitive to a subtle correlation between the numerical values 
of the Hamiltonian parameters -- a fact that is not common in the 
conventional cases of Anderson localization. Our results are analytically exact, 
and supported by numerical evaluation of the density of states and electronic 
transmission coefficient.
\end{abstract}
\maketitle
\section{Introduction}
\label{intro}
Electronic wave functions in a disordered lattice exhibit an exponentially localized 
envelope in space -- a phenomenon, commonly known as the Anderson localization
~\cite{anderson,kramer,abrahams,vollhardt}. The problem has kept itself alive and 
kicking over all these years in condensed matter physics, and has given quantum 
transport properties of disordered systems intriguing twists and turns. 
The recent development of fabrication and lithographic techniques has 
taken the phenomenon of Anderson localization beyond the electronic systems, 
substantiated by remarkable experiments incorporating localization of 
light~\cite{sperling,chen}, ultrasound in three dimensional elastic 
networks~\cite{hu}, or even plasmonic~\cite{tao,christ} and 
polaritonic~\cite{barinov,grochol} lattices. Direct observation of the 
localization of matter waves~\cite{damski,billy,roati,chabe,kondov} in recent times 
has made the decades old phenomenon even more exciting.

The key point in Anderson localization is the dimensionality. 
Within the tight binding approximation, the electronic wave functions are 
localized for dimensions $d \le 2$ (the band center in the off diagonal 
disorder case is an exception). For $d > 2$ with strong disorder,  
the wave function decays exponentially~\cite{kramer,abrahams}. 
Extensive analyses of the localization length~\cite{rudo1,rudo2}, 
density of states~\cite{alberto}, and multi-fractality 
of the single particles states~\cite{rudo3,rudo4} have consolidated 
the fundamental ideas of disorder induced localization. 
Intricacies of the single parameter scaling hypothesis -- its 
validity~\cite{rudo5}, variance~\cite{deych}, or even 
violation~\cite{bunde,titov} in low dimensional systems provided the  
finer details of the localization phenomenon that have subsequently 
been supported by experimental measurements 
of conductance distribution in quasi-one dimensional gold wires~\cite{mohanty}.

However, in low dimensions, or more specifically, in 
one dimensional disordered lattices even a complete delocalization of electronic 
states can be seen. This path breaking result was initially put forward by 
Dunlap et al.~\cite{dunlap} in connection with a sudden enhancement of 
conductance of a class of polyanilenes on protonation. Known as the 
{\it random dimer model} (RDM) the phenomenon is attributed to certain 
special kinds of positional correlation in the potential profiles. 
The investigation of delocalization of eigenstates in 
correlated disordered models was taken up further over the years and interesting 
results such as the relation of localization length with the density of 
states~\cite{trigan1} were put forward. The work extended to quasi-one dimensional 
systems as well for which the Landauer resistance and its relation to the localization 
length was examined in details~\cite{trigan2} for a two-leg ladder model, an extensive 
extension of which was later done by Sedrakyan et al.~\cite{trigan3}. 
Controlled disorder induced localization and delocalization of eigenfunctions took a 
considerable volume in contemporary literature, exploring solid non-trivial results 
involving electron or phonon eigenstates~\cite{francisco1,david,francisco2}. 
Extended eigenfunctions in all such works mostly appear at special 
discrete set of energy eigenvalues.  

Eventually, the possibility of a controlled engineering 
of spectral continuum populated by extended 
single particle states and even a metal-insulator transition in one, or 
quasi-one dimensional discrete systems have also been discussed in the 
literature~\cite{moura,maiti,rudo6}. But, on the whole, the general 
exponentially localized character of the eigenfunctions prevails, and 
the possibility of having a mixed spectrum of localized 
and extended states in a disordered system (under some special 
positional correlations) is now well established.

Can one generate, going beyond the RDM, a full band of only extended eigenfunctions 
in a disordered system with $d \le 2$ ? If yes, what would be the 
minimal models capable of showing such unusual spectra ?
This is the question that we address ourselves in the present communication.
We put forward examples of a class of essentially one 
dimensional disordered and quasiperiodic lattices where a {\it complete 
delocalization} of electronic states can be engineered,  and {\it absolutely 
continuous} bands can be formed in the energy spectrum.
This is shown to be possible when an infinite disordered or quasiperiodic array 
of two kinds of `bonds' is side coupled to a single or a cluster of quantum dots (QD)
from one side at a special set of vertices. Minimal requirements are discussed in details.
In some of the examples cited here, the attachment of the dots form local loops 
which can be pierced by a constant magnetic field, breaking the time reversal 
symmetry of electron-hopping only locally, along the edges of such closed loops.
The engineering of bands of extended states is shown to be the result of a 
definite numerical correlation in the values of the 
electron hopping amplitude along the chain (backbone) and 
the coupling of the linear backbone with the 
side coupled dots, the strength of the magnetic field or both.

It should be mentioned that an early report of a RDM-kind of correlation 
leading to extended eigenfunctions in a Fibonacci superlattice was put 
forward by Kumar and Ananthakrishna~\cite{kumar}. The insight into the 
phenomenon was immediately provided by Xie and Das Sarma~\cite{dassarma}. 
However, the fact that, certain specific {\it numerical relationship} among 
a subset of  parameters of the Hamiltonian is capable of producing, absolutely 
continuous {\it bands} of extended eigenfunctions is uncommon, and to the best 
of our knowledge, has not been addressed until very recently~\cite{biplabpal}. 

We consider two bonds $A$ and $B$ arranged along a line forming an infinite 
linear chain. The sequence of the bonds may be random or quasiperiodic~\cite{macia}, 
offering either a pure point spectrum or a singular continuous one. 
The bonds connect identical atomic sites, an infinite subset of which is 
coupled to similar atoms (mimicing single level quantum dots (QD)) from 
one side giving the system a quasi one dimensional flavor. The disorder 
(or, quasiperiodic order) thus has a topological character.  
In addition to the basic interest of going beyond the RDM, 
two other facts motivate us in undertaking such a work.

First, the Fano-Anderson effect~\cite{hopkins,mirosh1} caused by the 
insertion of a bound state into a continuum is an exciting field, and has been 
investigated recently in nanoscale systems~\cite{mirosh2}. In this context, 
our study provides examples where one can observe at least one effect of inserting 
multiple bound states, in fact, an infinity of them in a {\it singular continuum}, 
or a pure point spectrum.

Second, the present advanced stage of growth techniques has motivated in depth 
studies of quasiperiodic nanoparticle arrays in the context of ferromagnetic 
dipolar modes~\cite{negro1} or plasmon modes~\cite{negro2}. Also, the use of 
a scanning tunnel microscope (STM) tip to fabricate structures atom by atom, 
viz., Xe on Ni substrates~\cite{eigler}, or nanometer size gold particles on 
metals~\cite{mamin}, or, putting individual atoms of Si substrate~\cite{salling} 
has stimulated a lot of work in this field~\cite{vasseur1,vasseur2}. Our 
results can motivate future experiments in this direction.

In section~\ref{model} we describe the lattice models. 
In Section~\ref{method}, within subsections~\ref{method}A and \ref{method}B 
the local, non-local and the mixed cases introduced in 
section~\ref{model} are discussed, with explicit remarks on the density of states 
profiles in each case. Subsection~\ref{method}C specially deals with the 
special case of a Fibonacci quasiperiodic chain, using a real space renormalization 
group (RSRG) scheme. Section~\ref{transmission} describes the two terminal 
transmission coefficient, while section~\ref{univer} provides a critical 
discussion on the evolution of the parameter space under the RSRG scheme 
and its relation with the extendedness of the wave function. In 
section~\ref{othergeometries} we briefly point out a triplet of other 
geometries which are less restrictive compared to the ones discussed here, 
and in section~\ref{conclu} we draw our conclusion.

\section{The model}
\label{model}
\begin{figure}[ht]
\centering
\includegraphics[clip,width=8.5cm,angle=0]{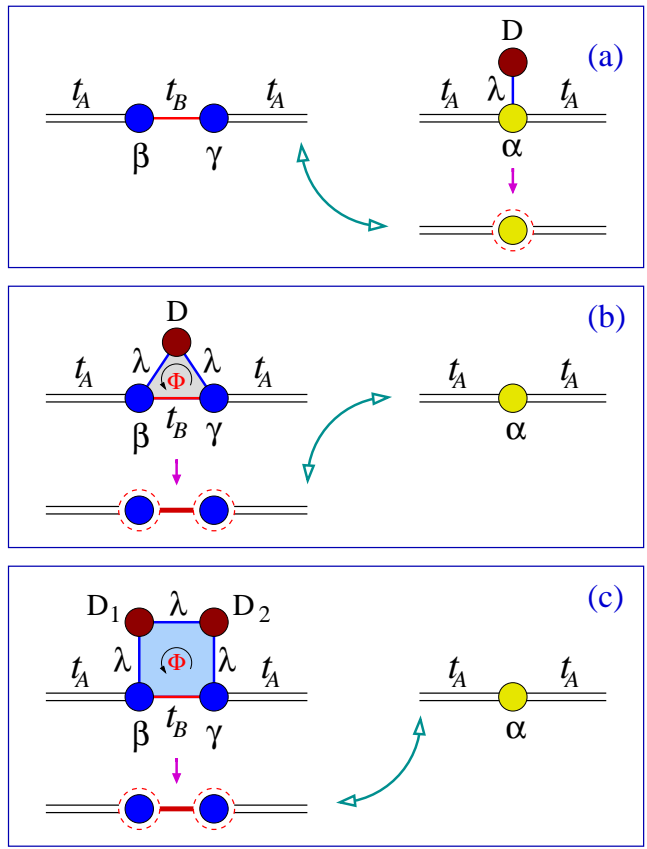}
\caption{(Color online) Building blocks of the quasi one-dimensional 
lattices described in the text. In each case the backbone is a linear 
array of two kinds of bonds $A$ (double line) and $B$ (red single line), 
such that a $B$-bond is always flanked by two $A$-bonds on either side. 
The atomic sites on the backbone are marked as $\alpha$, $\beta$ and $\gamma$ 
as described in the picture. The hooping integrals are appropriately 
described by $t_{A}$ and $t_{B}$. 
(a) A QD ($D$) is locally connected to the $\alpha$-site. This $D$-$\alpha$ 
cluster is ``renormalized" into an effective site (yellow circle 
surrounded by red dotted lines). 
(b)A QD ($D$) is non-locally coupled to the $\beta$-$\gamma$ pair. The 
$D$-$\beta$-$\gamma$ cluster is then renormalized into the immediate 
lower geometry, pointed by the arrow.  
(c)The QDs $D_{1}$ and $D_{2}$ exhibit a mixed connection to $\beta$-$\gamma$ 
pair. The block $\beta$-$D_{1}$-$D_{2}$-$\gamma$ is renormalized to the diatomic 
molecule shown by the arrowhead. 
In every case, the linear chain (disordered or quasiperiodic) is formed by 
arranging the cluster linked by the bent cyan double arrowheads in the desired 
order.} 
\label{cells}
\end{figure}

We refer the reader to Fig.~\ref{cells} 
where the basic structural units are 
displayed. The backbone in each case is an infinite array of a single (red) bond $B$ 
and a double bond $A$. We shall restrict ourselves to a geometry where the 
single `$B$' bonds do not come pairwise. Thus we have a kind of 
`anti-RDM' here. This is not always needed though, as will be discussed 
in the concluding section. 

Three cases are separately discussed. The simplest one is that of a local 
connection (LC), where a single QD (marked as $D$ in Fig.~\ref{cells}(a) 
is tunnel-coupled to a site $\alpha$ flanked by two $A$-bonds. The second 
case discusses a non-local connection (NLC), where a QD ($D$) is  
tunnel-coupled to both the sites residing at the extremities ($\beta$ and $\gamma$ in 
Fig.~\ref{cells}(b)) of a $B$-bond. The final geometry describes a mixed 
connection (MC), where two inter-coupled QDs $D_{1}$ and $D_{2}$ are connected to the 
extremities of a $B$-bond (i.e. to $\beta$ and $\gamma$ sites) as shown in 
Fig.~\ref{cells}(c). In the two latter cases a uniform magnetic field is 
applied in a direction  perpendicular to the plane of every closed loop. 
The system in each case is described by a tight-binding Hamiltonian. 

We show that, for a particular algebraic relationship between the nearest 
neighbor hopping integrals $t_{ij}$ along the backbone and the backbone-QD 
coupling $\lambda$, the infinite topologically disordered or quasiperiodic 
chain of scatterers yields absolutely continuous energy bands in the spectrum. 
In the case of LC (Fig.~\ref{cells}(a)) there will be two continuous subbands. 
In the NLC and MC cases (Fig.~\ref{cells}(b) and (c)) a single absolutely continuous 
band spans the entire energy spectrum when, in addition to the algebraic 
relationship between the hopping integrals $t_{ij}$ and $\lambda$, the 
magnetic flux $\Phi$ threading each elementary plaquette assumes a particular value. 
These two cases (NLC and MC) therefore represent situations where the 
spectral character can be grossly changed from pure point or singular 
continuous to absolutely continuous by tuning an external magnetic field. 
This may be useful from the standpoint of device technology.
\section{The Hamiltonian and the general scheme}
\label{method}
Spinless, non-interacting electrons on the chain comprising the building blocks 
depicted in Fig.~\ref{cells} are described by the Hamiltonian,  
\begin{equation}
{\bm H} = \epsilon \sum_{i} c_{i}^{\dagger} c_{i} + \sum_{\langle ij 
\rangle} t_{ij} \left[c_{i}^{\dagger} c_{j} + h.c. \right] 
\label{hamilton}
\end{equation}
where, $\epsilon$ is the constant on-site potential, at every site including 
the QD (marked $D$). We have colored the atomic sites differently just to 
distinguish between their nearest neighbor bond configurations. These are 
marked as $\alpha$ (yellow circle), $\beta$ and $\gamma$ (blue circles) 
respectively. The nearest-neighbor hopping integral $t_{ij} = t_{A}$ 
(double bond) along the backbone on either side of an $\alpha$-site, while 
it is $t_{B}$ (denoted by red line segment) between a $\beta$-$\gamma$ pair. 
In the LC case (Fig.~\ref{cells}(a)) $t_{ij} = \lambda$ between the QD and 
the $\alpha$-site. In the NLC and the MC situations (Fig.~\ref{cells}(b) 
and (c)) the presence of the magnetic flux breaks the time reversal symmetry 
along the edges of the loops. This is taken care of by incorporating the 
appropriate Peierls' phase factor in the hopping integrals, viz., $t_{ij} 
\rightarrow t_{ij} \exp{\theta_{ij}}$ where, 
$\theta_{ij}= 2\pi\Phi a_{\ij}/(L\Phi_{0})$. $L$ is the perimeter of the 
plaquette and $a_{ij}$ is the length of the bond connecting the $i$-th 
and the $j$-th sites of the loop. $\Phi_{0}=hc/e$ is the flux quantum.

Let us consider symmetric geometries only. This means that, in the 
NLC case, we assume that the QD is placed symmetrically above the 
$\beta$-$\gamma$ cluster. In the MC case similarly, the 
$\beta$-$D_{1}$, $D_{1}$-$D_{2}$ and the $D_{2}$-$\gamma$ distances are equal. 
This just simplifies the mathematical expressions without sacrificing any 
physics that we are going to establish. Thus, in the NLC (Fig.~\ref{cells}(b)) 
$t_{\beta\gamma}=t_{\gamma\beta}^{*}=t_{B} \exp{(i \theta_{1,NL})}$, 
$t_{\gamma D}=t_{D\beta}=t_{\beta D}^{*}=
t_{D \gamma}^{*}=\lambda \exp{(i\theta_{2,NL})}$, 
with $\theta_{1,NL}=2\pi\Phi a_{1}/(a_{1}+2a_{2})\Phi_{0}$ and 
$\theta_{2,NL}=2\pi\Phi a_{2}/(a_{1}+2a_{2})\Phi_{0}$,
$a_{1}$ and $a_{2}$ being the bond lengths between the $\beta$-$\gamma$ pair, 
and the $\beta$-$D$ and $\gamma$-$D$ pairs respectively. The asterisk denotes the 
complex conjugate.
   
Similarly, in the MC case (Fig.~\ref{cells}(c)), 
$t_{\beta\gamma}=t_{B} \exp{(i\theta_{1,M})}$, and  
$t_{\gamma D_{2}}=t_{D_{2} D_{1}}=t_{D_{1} \beta}=
\lambda \exp{(i\theta_{2,M})}$. In this case however,
$\theta_{1,M}=2\pi\Phi a_{1}/(a_{1}+3a_{2})\Phi_{0}$ and 
$\theta_{2,M}=2\pi\Phi a_{2}/(a_{1}+3a_{2})\Phi_{0}$ 
where, $a_1$ and $a_2$ are the bond lengths between the $\beta$-$\gamma$ 
pair, and the $\beta$-$D_{1}$, $D_{1}$-$D_{2}$ and $D_{2}$-$\gamma$ 
pairs respectively. The respective complex conjugates are trivially understood.

Using the difference equation version of the Schr\"{o}dinger equation, viz., 
\begin{equation}
(E - \epsilon) \psi_{i} = \sum_{j} t_{ij} \psi_{j}
\label{diff}
\end{equation}
we decimate out the vertices (QDs) in each of the three cases to map the local, 
non-local and mixed clusters on to effective atomic sites with renormalized 
on-site potentials given by, 
$\epsilon_{\alpha}=\epsilon+\lambda^{2}/(E-\epsilon)$ in the LC (Fig.~\ref{cells}(a)),
$\epsilon_{\beta}=\epsilon_{\gamma}=\epsilon+\lambda^{2}/(E-\epsilon)$ 
in the NLC (Fig.~\ref{cells}(b)), and 
$\epsilon_{\beta}=\epsilon_{\gamma}=\epsilon+\lambda^{2} (E-\epsilon)/\Delta$ 
in the MC case (Fig.~\ref{cells}(c)), where, 
$\Delta=(E-\epsilon)^{2}-\lambda^{2}$. 
The sites with renormalized on-site potential in each case are encircled with 
the red dotted lines in Fig.~\ref{cells}.
The hopping integrals are still $t_{A}$ and 
$t_{B}$ along the linear backbone for the LC, 
while they are, $t_{\beta\gamma}=t_{\gamma\beta}^{*}=
t_{B} \exp{(i\theta_{1,NL})}+\lambda^{2} \exp{(-2i\theta_{2,NL})}/(E-\epsilon)$ 
in the NLC case, and 
$t_{\beta\gamma}=t_{\gamma\beta}^{*}=t_{B} \exp{(i\theta_{1,M})}+
\lambda^{3} \exp{(-3i\theta_{2,M})}/\Delta$ in the MC case.

One can now build up an infinite chain of $\alpha$ sites (renormalized, 
in the LC case) and the $\beta$-$\gamma$ doublet (renormalized in the NLC 
and MC cases) in any desired order. The amplitude of the wave function at 
any remote site on such a chain is conveniently obtained by the transfer 
matrix technique. Using the difference equation the amplitudes of the wave 
function at the neighboring sites along the effective one dimensional chain 
can be related using the $2 \times 2$ transfer matrices, 
\begin{eqnarray}
\left(\begin{array}{c}
\psi_{n+1} \\
\psi_{n}  
\end{array} \right)
& = & 
\left(\begin{array}{cccc}
\dfrac{E-\epsilon_n}{t_{n,n+1}} & -\dfrac{t_{n,n-1}}{t_{n,n+1}} \\ 
1 & 0 
\end{array} 
\right)
\left(\begin{array}{c}
\psi_{n} \\
\psi_{n-1} 
\end{array} \right)
\end{eqnarray}
The hopping integrals $t_{n,n\pm 1}$ will carry the appropriate 
phase factors when written for the NLC and MC cases.

It is obvious that there are three kinds of transfer matrices, viz., 
${\bm{M_{\alpha}}}$, ${\bm{M_{\beta}}}$, ${\bm{M_{\gamma}}}$ and 
which will differ in their matrix elements, depending on the 
respective on-site potentials and the nearest-neighbor hopping integrals. 
From the arrangement of the $\beta$-$\gamma$ clusters and the isolated sites in 
the original chain it can be appreciated that the the wave function at a far 
end of the chain can be determined if one evaluates the product of the 
unimodular matrices ${\bm{M_{\alpha}}}$ and 
${\bm{M_{\gamma \beta}}} = {\bm{M_{\gamma}}}.{\bm{M_{\beta}}}$  
sequenced in the desired random or quasiperiodic fashion.

The central result of this communication is that, in each of the three cases 
of LC, NLC and MC, the commutator $[{\bm{M_{\alpha}}},{\bm{M_{\beta\gamma}}}]$ 
can be made to vanish irrespective of the energy $E$ of the electron whenever 
the system parameters are inter-related in a certain algebraic fashion. Let us 
look at the explicit expressions. We list below only one off diagonal element of 
the commutator for every configuration (LC, NLC or MC), as the diagonal elements 
of the above commutator vanish identically in each case, and 
$[{\bm{M_{\alpha}}},{\bm{M_{\gamma\beta}}}]_{21}=
[{\bm{M_{\alpha}}},{\bm{M_{\gamma\beta}}}]_{12}$.
\begin{itemize}
\item {\it The Local Coupling:} In this case, 
\begin{equation}
[{\bm{M_{\alpha}}},{\bm{M_{\gamma \beta}}}]_{12} = 
\dfrac{\lambda^2-(t_{B}^{2}-t_{A}^{2})}{t_{A} t_{B}}
\label{lc}
\end{equation}
\item {\it The Non Local Coupling:} Here, 
\begin{eqnarray}
&[{\bm{M_{\alpha}}},{\bm{M_{\gamma\beta}}}]_{12} = \nonumber \\ 
&\dfrac{(E-\epsilon) e^{i 2\pi\Phi/\Phi_{0}} (t_{A}^{2}-t_{B}^{2}-\lambda^{2}) - 
2 \lambda^{2} t_{B} \cos (\dfrac{2\pi\Phi}{\Phi_{0}})}  
{t_{A} e^{i\theta_{1,NL}} [\lambda^{2} + (E-\epsilon)t_{B} e^{i 2\pi\Phi/\Phi_{0}}]} 
\label{nlc}
\end{eqnarray}
and,
\item {\it The Mixed Coupling:}
In this case,
\begin{eqnarray}
&[{\bm{M_{\alpha}}},{\bm{M_{\gamma\beta}}}]_{12} = \nonumber\\
&\dfrac{e^{i 2\pi\Phi/\Phi_{0}} [(E-\epsilon)^{2}-\lambda^{2}] 
(t_{A}^{2} - t_{B}^{2} - \lambda^{2})
- 2t_{B} \lambda^{3} \cos (\dfrac{2\pi\Phi}{\Phi_{0}})}
{t_{A} e^{i\theta_{1,M}} \left[t_{B} e^{i 2\pi\Phi/\Phi_{0}} 
[(E-\epsilon)^{2}-\lambda^{2}] + \lambda^{3} \right ]}  
\label{mc}
\end{eqnarray}
\end{itemize}
A look at Eqs.~\eqref{lc}-\eqref{mc} reveals that it is possible to make the 
commutator vanish independent of the energy $E$. Let us discuss case by case.
\subsection{The local coupling}
Eq.~\eqref{lc} shows that $[{\bm{M_{\alpha}}},{\bm{M_{\gamma\beta}}}]_{12}$ 
(and hence the full commutator) vanishes if we set  
\begin{equation}
\lambda = \pm \sqrt{t_{B}^{2} - t_{A}^{2}}
\end{equation}
This implies that, with the above value of the tunnel hopping integral, the 
electronic energy spectrum will no longer be sensitive to the arrangement of 
the matrices ${\bm{M_{\alpha}}}$ and ${\bm{M_{\gamma\beta}}}$, 
that is, independent of the arrangement of the atomic site $\alpha$, and 
the pair $\beta$-$\gamma$. This happens independent of the energy $E$ of the 
electron. This result needs to be contrasted clearly with that in the 
RDM~\cite{dunlap} where the local structure of disorder could transform one 
subset of the transfer matrices into unit matrices, but only at special 
value of $E$. In our case, with the commutation condition satisfied one can 
arrange the constituent elements $\alpha$ and $\beta$-$\gamma$ even in any kind 
of perfect periodic order. The wave functions as a result, will have to be 
of a {\it perfectly extended}, Bloch-like character and the energy bands 
will exhibit absolutely continuous measure whenever 
$\lambda=\pm\sqrt{t_{B}^{2} - t_{A}^{2}}$. However, this condition is only 
necessary, and we discuss below the {\it sufficient} condition for observing 
extended eigenstates.

\begin{figure}[ht]
\includegraphics[clip,width=8.5cm,angle=0]{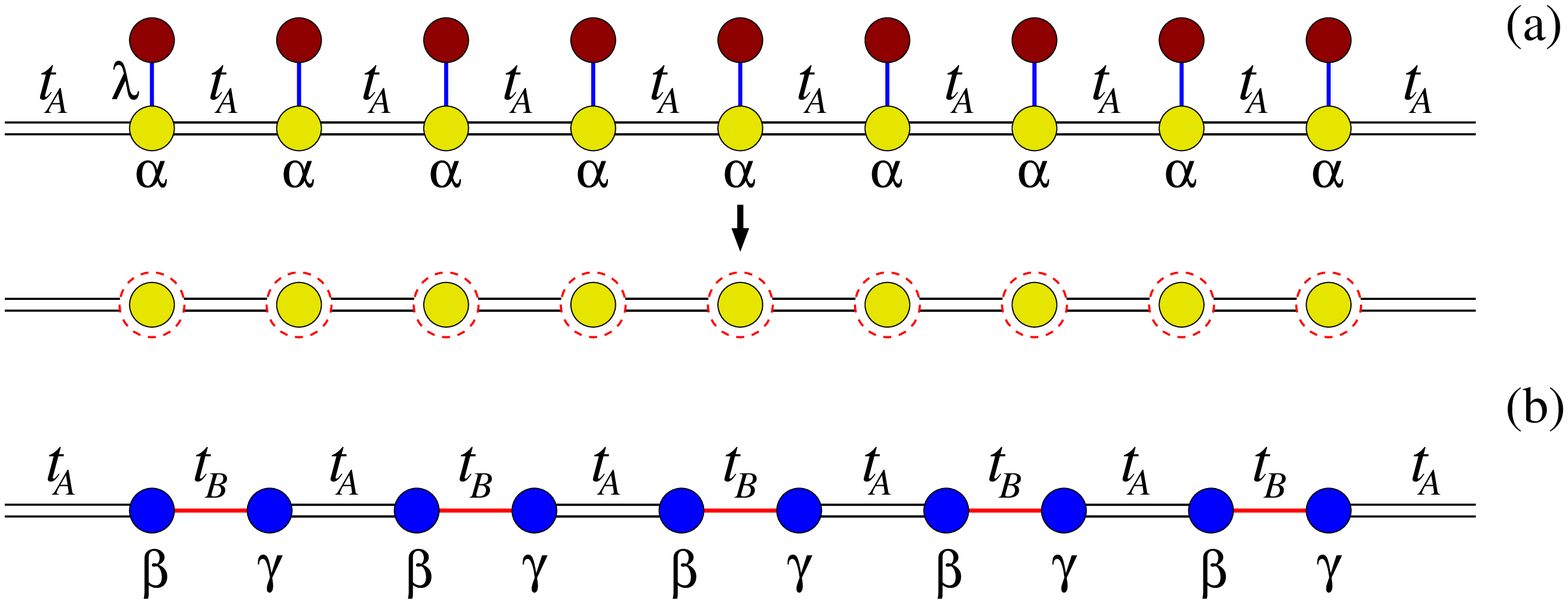}
\caption{(Color online) (a) Periodic arrangement of $\alpha$ sites (yellow) 
coupled to the QD (red) and its renormalized version where the renormalized 
$\alpha$ sites are encircled by dotted red lines, and (b) periodic array 
of $\beta$-$\gamma$ pairs. The strength of the hopping $t_{B}$ (red line) is taken 
to be greater than $t_{A}$ (double lines).} 
\label{order}
\end{figure}

Taking advantage of the commutation of the transfer matrices we can shuffle any 
arrangement of the atoms into two infinite, periodic arrays of the effective 
renormalized $D$-$\alpha$ cluster and $\beta$-$\gamma$ clusters (Fig.~\ref{order}).
The local density of states (LDOS) at 
any site of these lattices can be worked out analytically, and for the $\alpha$
 and $\beta$ sites the results are,
\begin{equation}
\begin{aligned}
\rho_{\alpha} =& \dfrac{1}{\pi} 
\dfrac{E - \epsilon}{\sqrt{4t_{A}^{2} (E-\epsilon)^{2} - 
[(E-\epsilon)^{2} - \lambda^{2}]^{2}}}\\
\rho_{\beta} =& \dfrac{1}{\pi} 
\dfrac{E - \epsilon}{\sqrt{4t_{A}^{2} (E-\epsilon)^{2} - 
[(E-\epsilon)^{2} - (t_{B}^{2} - t_{A}^{2})]^{2}}}
\end{aligned}
\label{merge1}
\end{equation}
In each case, the LDOS exhibits a continuous two-subband structure (typical 
of a one dimensional binary ordered chain). It is obvious that, with the 
{\it resonance} condition $\lambda=\pm \sqrt{t_{B}^{2} - t_{A}^{2}}$ the 
LDOS in the two cases overlap. That is the bands formed by each individual 
periodic sublattices merge completely. So, a linear array of  the structural 
units $\alpha$-$D$ and the $\beta$-$\gamma$ clusters, grown following any chosen 
pattern (for example, completely disordered, or quasiperiodic geometry) 
should also exhibit precisely these absolutely continuous sunbands. As 
extended and localized eigenstates can not coexist at the same energy, 
the electronic states must be of an {\it extended character}, a fact that 
is substantiated later by a flow of the hopping integrals under RSRG and 
a perfect two terminal transmission. This completes the proof that in the 
LC case, a suitable choice of the hopping integrals can generate absolutely 
continuous subbands populated only with {\it extended} single particle states. 
\subsection{The non local and the mixed coupling}
We now turn our attention to the cases of NLC and MC which essentially refer 
to an array of triangle shaped and square plaquettes threaded by a magnetic 
flux and single atomic sites (Fig.~\ref{cells}(b) and (c)). The matrix 
elements $[{\bm{M_{\alpha}}},{\bm{M_{\gamma\beta}}}]_{12}$, 
as given by Eqs.~\eqref{nlc} and \eqref{mc} become zero 
in either situation when, $\lambda = \pm \sqrt{t_{A}^{2} - t_{B}^{2}}$, 
and, in addition to it, $\Phi = \Phi_{0}/4$ in either case.  
It means that, even if we fix $\lambda = \pm \sqrt{t_{A}^{2} - t_{B}^{2}}$ 
at the very outset, we still need to tune the magnetic flux $\Phi$ through 
each plaquette to a particular value to have 
$[{\bm{M_{\alpha}}},{\bm{M_{\gamma\beta}}}]=0$ {\it independent of the 
energy $E$ of the electron}. Just as before, we can now, using the 
commutivity of ${\bm{M_{\alpha}}}$ and ${\bm{M_{\gamma\beta}}}$ 
shuffle the building blocks to generate two infinite periodic chains 
corresponding to both the NLC and the MC cases, comprising of $\beta$-$\gamma$ 
pairs, and isolated single sites $\alpha$ ($\epsilon_{\alpha}=\epsilon$, in 
both these cases). In terms of the parent lattices, in the NLC situation this 
means that the single $\alpha$ sites and the $\beta$-$D$-$\gamma$ triangle can 
be arranged in any desired pattern, while for the MC case its any arbitrary 
linear arrangement of the $\alpha$ and the $\beta$-$D_{1}$-$D_{2}$-$\gamma$ cluster.
\begin{figure*}[ht]
\centering
\includegraphics[clip,width=18cm,angle=0]{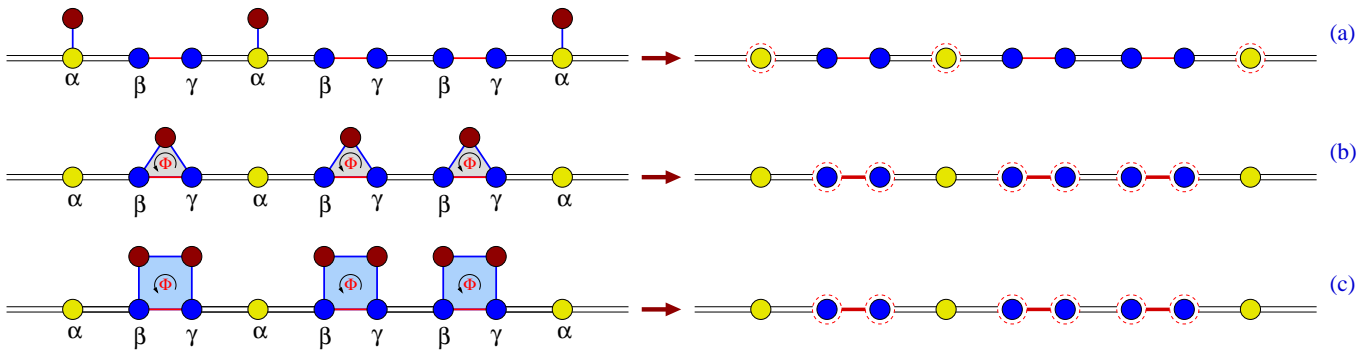}
\caption{(Color online) Lattices with the structural units placed in quasiperiodic  
Fibonacci order along the principal axis (backbone). (a) Local coupling with 
a QD attached to every $\alpha$-sites. (b) The non-local coupling and (c)
the mixed coupling cases. The linear chains with the renormalized $\alpha$-sites 
and the $\beta$-$\gamma$ clusters (or with $\alpha$-sites and renormalized $\beta$-$\gamma$ 
doublets), obtained by decimating the QDs (red circle) in every cases are shown for 
the LC, NLC and MC cases on the right in (a), (b) and (c) respectively. }  
\label{lattices}
\end{figure*}

The $\alpha$-lattice has the well known density of states, 
viz., $\rho_\alpha = (1/\pi)[4t_{A}^{2} - (E-\epsilon)^{2}]^{-1/2}$. To make 
things look algebraically simple, let us set $\lambda=t_{B}$, which just means 
that the side coupled QD is equispaced from the base sites, and that the phase 
acquired by the electron while hopping along an arm of a triangle as well as 
of a square is same for all the arms. The resonance condition now boils down 
to $\lambda = t_{B} = t_{A}/\sqrt{2}$ and of course, $\Phi = \Phi_{0}/4$. 
The LDOS at the $\beta$ site corresponding to the NLC case is given by, 
$\rho_{\beta}^{NLC} = (1/\pi) (\mathcal{F}_{NLC})^{-1/2}$ where, 
\begin{widetext}
\begin{equation}
\mathcal{F}_{NLC} = \dfrac{4 t_{A}^{2} t_{B}^{2} [(E-\epsilon)^{2} + 
2 (E-\epsilon) t_{B} \cos (2\pi\Phi/\Phi_{0}) + t_{B}^2]}
{[(E-\epsilon)^{2} - t_{B}^{2}]^{2}} - 
\left[E - \dfrac{\epsilon (E-\epsilon)^{2} + 2t_{B}^{3} \cos(2\pi\Phi/\Phi_0)
+ t_{A}^{2} (E-\epsilon) + t_{B}^{2} (2 E - 3 \epsilon)}  
{(E-\epsilon)^{2} - t_{B}^{2}}\right]^{2}
\end{equation}
\end{widetext}
and, the same corresponding to the MC case is given by 
$\rho_{\beta}^{MC} = (1/\pi) (\mathcal{F}_{MC})^{-1/2}$
where,
\begin{equation}
\mathcal{F}_{MC} = \xi_{1}(E,\epsilon,t_{A},t_{B},\Phi) + 
\xi_{2}(E,\epsilon,t_{A},t_{B},\Phi)
\end{equation}  
$\xi_{1}$ and $\xi_{2}$ are given by,
\begin{widetext}
\begin{equation}
\begin{aligned}
\xi_{1}(E,\epsilon,t_{A},t_{B},\theta) &= 
\dfrac{4t_{A}^{2} t_{B}^{2} \left[(E-\epsilon)^{4} - 
4t_{B}^{2}[(E-\epsilon)^{2}-2t_{B}^{2}] \sin^{2}(\pi\Phi/\Phi_{0})\right]}
{(E-\epsilon)^{2}[(E-\epsilon)^{2}-2t_{B}^{2}]^2}\\
\xi_{2}(E,\epsilon,t_{A},t_{B},\theta)&= \left[
\dfrac{(E-\epsilon)(\delta-t_{B}^{2})}{\delta}-
\dfrac{\delta^2(t_{A}^{2}+t_{B}^{2}) + 
t_{B}^{4} (t_{B}^{2}+2\delta \cos(2\pi\Phi/\Phi_{0}))}
{\delta (E-\epsilon)(\delta - t_{B}^{2})}\right]
\end{aligned}
\end{equation}
\end{widetext}
with $\delta = (E-\epsilon)^{2} - t_{B}^{2}$.

It is interesting to note that the algebraic expressions in the NLC and MC 
cases reduce to the simple form 
$\rho_{\alpha} = (1/\pi)[4t_{A}^2-(E-\epsilon)^{2}]^{-1/2}$ as soon as we 
set $\lambda = t_{B} = t_{A}/\sqrt{2}$ and $\Phi = \Phi_{0}/4$. This happens 
to be the LDOS at the $\alpha$-site of a pure $\alpha$-chain. The band extends 
from $E=\epsilon - 2t_{A}$ to $E=\epsilon + 2t_{A}$. Thus, the same resonance 
condition, viz., $\lambda=t_{B}=t_{A}/\sqrt{2}$ and $\Phi=\Phi_{0}/4$ results 
in a complete overlap of the energy bands at least in the energy range 
$[\epsilon-2t_A,\epsilon+2t_A]$ in both the cases. We have a single 
absolutely continuous band of extended eigenfunctions.
\subsection{Quasiperiodic Fibonacci order} 
As a specific example, we explicitly calculate the LDOS at the $\beta$-sites in a 
golden mean Fibonacci quasiperiodic chain. The chain is grown recursively following 
the usual Fibonacci inflation rule $A \rightarrow AB$ and $B \rightarrow A$
~\cite{macia}. The 
corresponding hopping integrals $t_{A}$ and $t_{B}$ follow a Fibonacci arrangement. The 
local, non-local or the mixed attachments of the QDs are shown in Fig.~\ref{lattices}. 
The `quasi one-dimensionality' caused by the side coupled clusters are removed by 
decimating the attachments and creating an {\it effective} one dimensional chain in 
each case, as depicted in the same figure. The decimation results in renormalized 
values of the on-site potentials at the $\alpha$-site in the LC case, and at the 
$\beta$, and the $\gamma$-sites in the NLC and the MC cases, as already mentioned.

Such a quasiperiodic Fibonacci chain is, by construction, self similar and allows an exact 
implementation of the RSRG methods. Renormalized versions of the Fibonacci chain are obtained 
by the well known decimation scheme~\cite{pal}. For the sake of understanding and to facilitate 
a subsequent discussion on the flow in parameter space we present the explicit RSRG recursion 
relations connecting the $(n+1)$-th and the $n$-th stages of iteration for the three cases.
\begin{itemize}
\item[$\blacklozenge$] {\it The Local Connection}:
\begin{equation}
\begin{aligned}
&\epsilon_{\alpha,n+1}  =  \epsilon_{\gamma,n} + 
\dfrac{t_{A,n}^{2} + t_{B,n}^{2}}{E - \epsilon_{\beta,n}}\\
&\epsilon_{\beta,n+1}  =  \epsilon_{\gamma,n} + 
\dfrac{t_{B,n}^{2}}{E - \epsilon_{\beta,n}}\\
&\epsilon_{\gamma,n+1}  =  \epsilon_{\alpha,n} + 
\dfrac{t_{A,n}^{2}}{E - \epsilon_{\beta,n}}\\
&t_{A,n+1}  =  \dfrac{t_{A,n} t_{B,n}}{E - \epsilon_{\beta,n}}\\
&t_{B,n+1}  =  t_{A,n}
\end{aligned} 
\label{rgrecur1}
\end{equation}
\end{itemize}
with, $\epsilon_{\alpha,0}=\epsilon + 
\lambda^{2}/(E-\epsilon)$, $\epsilon_{\beta,0}=\epsilon_{\gamma,0}=\epsilon$, 
$t_{A,0} = t_{A}$ and $t_{B,0}=t_{B}$.
\begin{itemize} 
\item[$\blacklozenge$] {\it The Non-Local and the Mixed Coupling}:
\end{itemize}

In both these cases, the magnetic flux breaks the time reversal symmetry, but only 
locally, along the $B$ bonds connecting the $\beta$-$\gamma$ vertices of the linear 
chain in the right panels of 
Fig.~\ref{lattices}(b) and (c). For this we designate by $t_B^f$ and $t_B^b$ 
the {\it forward} and {\it backward} hopping respectively along the $B$ bond. This 
naturally takes care of the phase introduced by the field along this segment. The 
hopping $t_A$ along the $A$ bond, though free from any phase at the bare length 
scale, picks up phase on renormalization which needs to be taken care of. The 
recursion relations for both the chains are, 
\begin{equation}
\begin{aligned}
&\epsilon_{\alpha,n+1}  =  \epsilon_{\gamma,n} + 
\dfrac{t_{A,n}^{f} t_{A,n}^{b} + t_{B,n}^{f} t_{B,n}^{b}}{E - \epsilon_{\beta,n}}\\
&\epsilon_{\beta,n+1}  =  \epsilon_{\gamma,n} + 
\dfrac{t_{B,n}^{f} t_{B,n}^{b}}{E - \epsilon_{\beta,n}}\\
&\epsilon_{\gamma,n+1}  =  \epsilon_{\alpha,n} + 
\dfrac{t_{A,n}^{f} t_{A,n}^{b}}{E - \epsilon_{\beta,n}}\\
&t_{A,n+1}^{f}  =  \dfrac{t_{A,n}^{f} t_{B,n}^{f}}{E - \epsilon_{\beta,n}}\\
&t_{B,n+1}^{f}  =  t_{A,n}^{f}
\end{aligned} 
\label{rgrecur2}
\end{equation}
The complex conjugate hopping integrals are defined appropriately. The initial 
values are of course different in these two cases, and are given by, 
$\epsilon_{\alpha,0}=\epsilon$, $\epsilon_{\beta,0}=\epsilon_{\gamma,0}=
\epsilon+\lambda^{2}/(E-\epsilon)$; $t_{A,0}^{f}=(t_{A,0}^{b})^*=t_{A}$ and 
$t_{B,0}^{f} = (t_{B,0}^{b})^*=t_{B}\exp(i\theta) + \lambda^{2}\exp(-2i\theta)/(E-\epsilon)$ 
in the NLC case, while, 
$\epsilon_{\alpha,0}=\epsilon$, $\epsilon_{\beta,0}=\epsilon_{\gamma,0}=
\epsilon+\lambda^{2}(E-\epsilon)/\Delta$; $t_{A,0}^{f}=(t_{A,0}^{b})^*=t_{A}$ and 
$t_{B,0}^{f}=(t_{B,0}^{b})^*=t_{B}\exp(i\theta) + \lambda^{3}\exp(-3i\theta)/\Delta$ and 
$\Delta=(E-\epsilon)^{2}-\lambda^{2}$ in the MC case. 
The phase $\theta=2\pi\Phi/3\Phi_0$ in the NLC case and it is  
$\theta=2\pi\Phi/4\Phi_0$ in the MC one. 
\begin{figure}[ht]
\centering
\includegraphics[clip,width=7cm,angle=0]{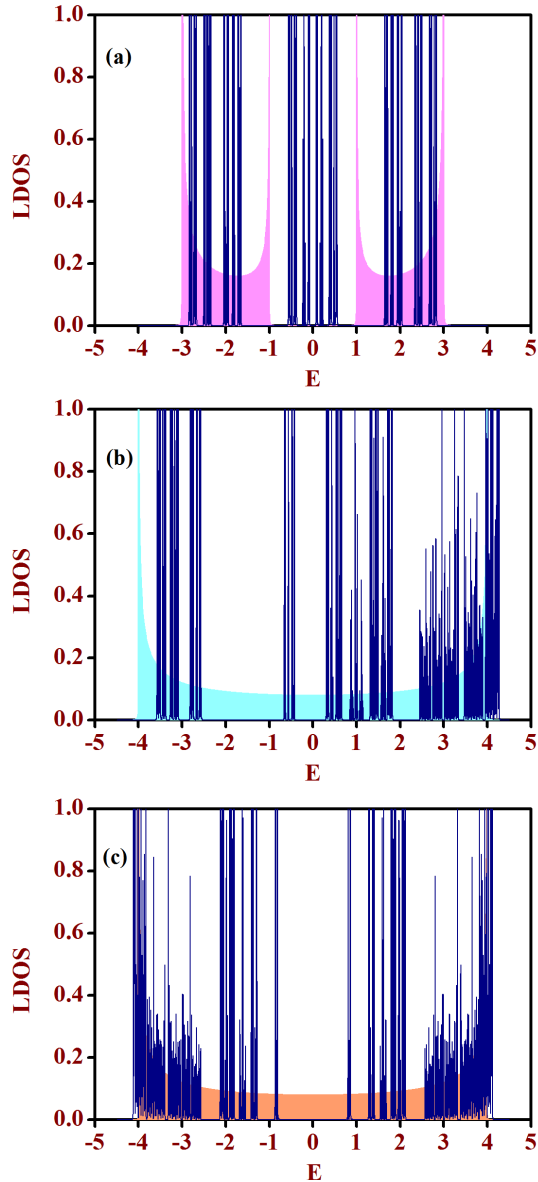}
\caption{(Color online) Local density of states (LDOS) at the $\beta$ site 
of an infinite Fibonacci array for (a) the locally connected QDs, 
(b) a single QD non-locally connected to every $\beta$-$\gamma$ pair, and 
(c) the mixed case of directly and indirectly coupled QDs to the 
$\beta$-$\gamma$ pair. In each panel, the fragmented display represents the 
off-resonance case while the absolutely continuous sub-bands or band 
represent the cases when $[{\bm{M_{\alpha}}},{\bm{M_{\gamma\beta}}}]=0$.
We have set $\epsilon=0$ in all the cases. $t_A=1$ and $t_B=2$ in (a) while 
$t_A=1$ and $t_B=t_A/\sqrt{2}$ in (b) and (c).}
\label{density}
\end{figure}
At every stage of renormalization the renormalized {\it forward} and 
{\it backward} hopping integrals are, of course, complex conjugate of each other.

The local Green's function at any $j$-th site ($j=\alpha, \beta$ or $\gamma$) 
is given by $G_{00}=(E-\epsilon_{j}^{*})^{-1}$ where, $\epsilon_{j}^{*}$ is the 
fixed point value of the corresponding on-site potential obtained by repeated 
application of the set of Eq.~\eqref{rgrecur1} and Eq.~\eqref{rgrecur2} for 
the local or the non-local and the mixed cases respectively. 
The LDOS $\rho_{j}$ is obtained from the standard formula 
$\rho_j=(-1/\pi)\text{ Im}[G_{00}(E+i\eta)]$ in the limit $\eta \rightarrow 0$. 
We present the results in Fig.~\ref{density}. 

In the top panel, the case of LC is shown. The LDOS is obtained at a $\beta$-site. 
The off-resonance case is characterized by the sharp fragmented LDOS profile that 
brings out the typical multifractal character of the wave functions in a quasiperiodic 
geometry. As the `resonance condition' $\lambda=\sqrt{t_{B}^{2}-t_{A}^{2}}$ (with 
$t_{B}>t_{A}$) is satisfied, the fragmented spectrum turns into two absolutely 
continuous subbands. 

In the middle and the bottom panels the continuous band in the NLC and MC cases are 
illustrated by the shaded area. Here we select $t_{A}>t_{B}$. The resonance condition 
in either case is obtained by setting $\lambda=t_{B}=t_{A}/\sqrt{2}$ and $\Phi=\Phi_{0}/4$. 
Deviating away from this generates the characteristic fragmented spectral form of a 
Fibonacci chain, as shown by the sharp blue lines (for $\Phi=0$) in each figure. The 
interesting difference with the LC case here is the existence of a single continuous 
band of states which will later be proven as extended, as shown by the shaded colored regions.

It should be appreciated that our purpose has been only to demonstrate the appearance 
of absolutely continuous part(s) in the energy spectrum. The LDOS coming from any one 
kind of sites is enough for this purpose. The contribution to the full density of states 
coming from the side-coupled QD sites generally consists of delta like localized peaks 
some of which reside outside the continuum~\cite{samar}. These are of no concern in 
the present discussion, as the central motivation has always been to prove the generation 
of a band of extended states only as a result of some algebraic correlation between the 
numerical values of the parameters of the Hamiltonian. The {\it extended} character of 
the eigenstates populating such continuous portions of the energy spectrum will 
subsequently be discussed in next sections.
\section{Transmission coefficient}
\label{transmission}
\begin{figure}[ht]
\centering
\includegraphics[clip,width=8.5cm,angle=0]{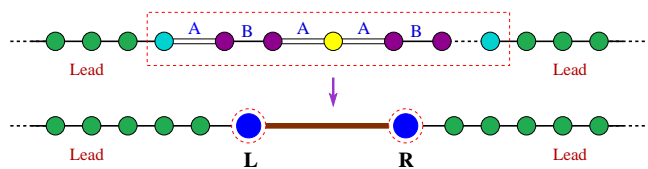}
\caption{(Color online) Geometry for calculation of the transmission 
coefficient. A Fibonacci array of bonds (enclosed in red dashed box) 
is clamped between two semi-infinite leads (green atoms), which is 
subsequently renormalized to a {\it dimer}, shown below by the blue 
atoms encircled by dotted red lines.}
\label{transport}
\end{figure}

To substantiate the LDOS profiles we also calculate the two terminal transport in the 
systems considered. The procedure is standard. The system is clamped between two perfectly 
periodic, semi-infinite leads on either side (Fig.~\ref{transport}). 
The sample trapped 
in between the leads is then decimated to a dimer by judiciously using the RSRG recurrence 
relations. Finally, the transmission coefficient is obtained by the well known 
formula~\cite{stone},
\begin{eqnarray}
&T=\dfrac{4\sin^{2}ka}{|\mathcal{A}|^{2}+|\mathcal{B}|^{2}} \\
&\text{with,}\quad \mathcal{A}=[(P_{12}-P_{21})+(P_{11}-P_{22})\cos ka] \nonumber\\
&\text{and}\quad \mathcal{B}=[(P_{11}+P_{22})\sin ka]\nonumber
\end{eqnarray}
where, $P_{ij}$ refer to the dimer-matrix elements, written appropriately in terms 
of the on-site potentials of the final renormalized {\it left} (L) and {\it right} (R) 
atoms $\epsilon_{L}$ and $\epsilon_{R}$ respectively, and the renormalized hopping 
between them~\cite{samar}. $\cos(ka)=(E-\epsilon_{0})/2t_{0}$, $\epsilon_{0}$ and 
$t_{0}$ being the on-site potential and the hopping integral in the leads, and $a$ 
is the lattice constant in the leads which taken equal to unity throughout the calculation.

\begin{figure}[ht]
\centering
\includegraphics[clip,width=7cm,angle=0]{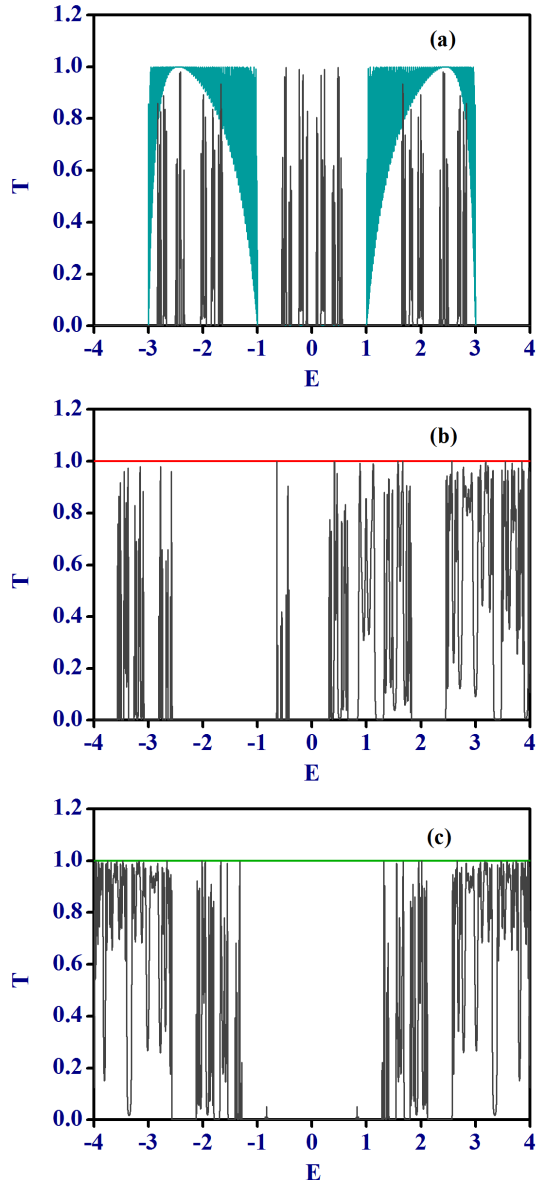}
\caption{(Color online) The variation of the transmission coefficient $T$ as a 
function of the energy $E$ of the electron for both the resonance and the 
off-resonance conditions. (a) represents the LC case, (b) represents the NLC case 
and (c) is for the MC case. The numerical values of the potentials and the 
hopping integrals are the same as in the LDOS figures.}
\label{trans}
\end{figure}

In Fig.~\ref{trans} we plot the transmission coefficient as a function of the energy 
of the electron in the three cases discussed so far. 
In each panel, again the resonance 
and off-resonance cases are plotted together for comparison. In the top panel, for the 
local coupling, when we set $\lambda=\sqrt{t_{B}^{2}-t_{A}^{2}}$, the transmission 
coefficient attains very high values, achieving the limit unity in most cases for the 
entire regions of the continuous subbands. There is a clean gap between the two zones 
of high transmittivity. It is because one has gaps in the energy spectrum in this region, 
and any {\it gap states} arising out of the side coupled dots in this part must have a 
localized character. The perfect transmission under the resonance condition in the LC 
case brings out a variation over the recent studies of Farchioni et al.~\cite{grosso}, 
where it was rightly shown that, side-coupled dots in general suppress the transmission 
across a linear tight binding chain.    

In the central and the bottom panels, the energy spectrum exhibits a single continuous 
band spanning the entire energy range. To be consistent with the LDOS figures we have 
preset $\lambda=t_{B}=t_{A}/\sqrt{2}$. The resonance, or a deviation from resonance is 
now controlled only by controlling the external magnetic field only. When the flux is 
detuned from its resonance value, the spectrum represents a fragmented character typical 
of quasiperiodic lattices, while precisely at $\Phi=\Phi_{0}/4$ the transmission coefficient 
turns out to be unity for the entire range of the continuum confirming the extended 
character of the eigenstates.
\section{RSRG flow pattern and extendedness of the eigenstates}
\label{univer}
In this section we would like to draw the attention of the reader to an interesting 
flow pattern followed by the on-site potentials and the hopping integrals when the 
elemental building blocks are arranged in a quasiperiodic Fibonacci chain, as discussed below.
\begin{figure}[ht]
\centering
\includegraphics[clip,width=8.5cm,angle=0]{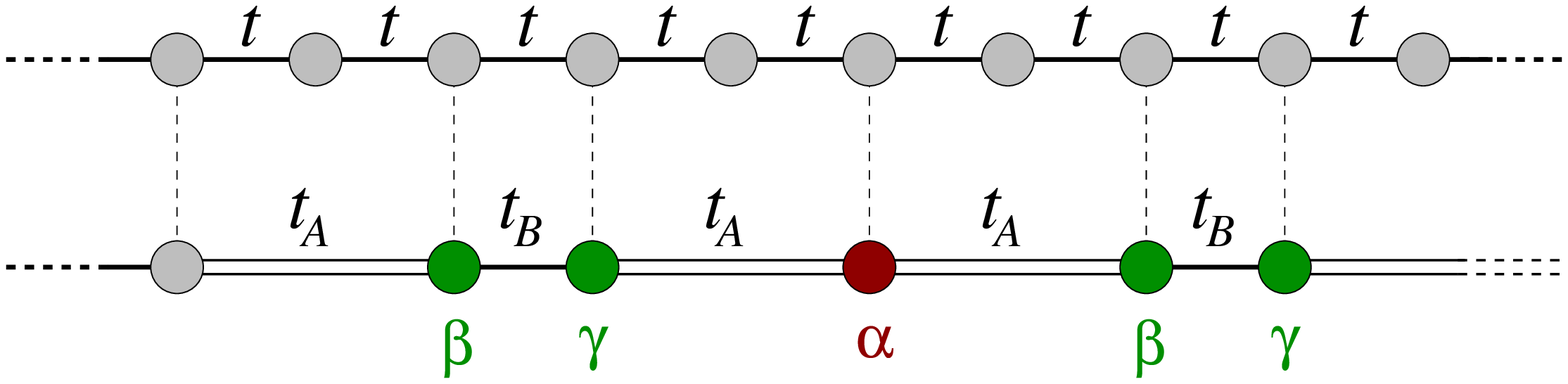}
\caption{(Color online) A periodically ordered chain, and a golden mean 
Fibonacci chain {\it artificially} generated from it  by a selective 
decimation of sites. The three kinds of sites in the lower chain 
are $\alpha$ (red), and $\beta$ and $\gamma$ (green), and the two nearest 
neighbor hopping integrals are $t_{A}$ and $t_{B}$ respectively.}
\label{rgflow}
\end{figure}

First, it should be noted that, since the transfer matrices ${\bm{M_{\alpha}}}$ 
and ${\bm{M_{\gamma\beta}}}$ corresponding to the the structural units depicted in 
Fig.~\ref{cells}(a)-(c) commute independent of energy $E$ under the appropriate {\it resonance} 
condition, the energy spectrum in this case should be the same for any uncorrelated disordered 
or quasiperiodic chains. As far as the quasiperiodic chains are concerned, though we have 
discussed the results specifically in terms of the golden mean Fibonacci sequence, the idea 
and subsequent results hold true for any generalized Fibonacci chain grown following the rule 
$A \rightarrow A^{n}B$ and $B \rightarrow A$, $A$ and $B$ representing the two bonds and $n \ge 1$.
\begin{figure*}[ht]
\centering
\includegraphics[clip,width=13cm,angle=0]{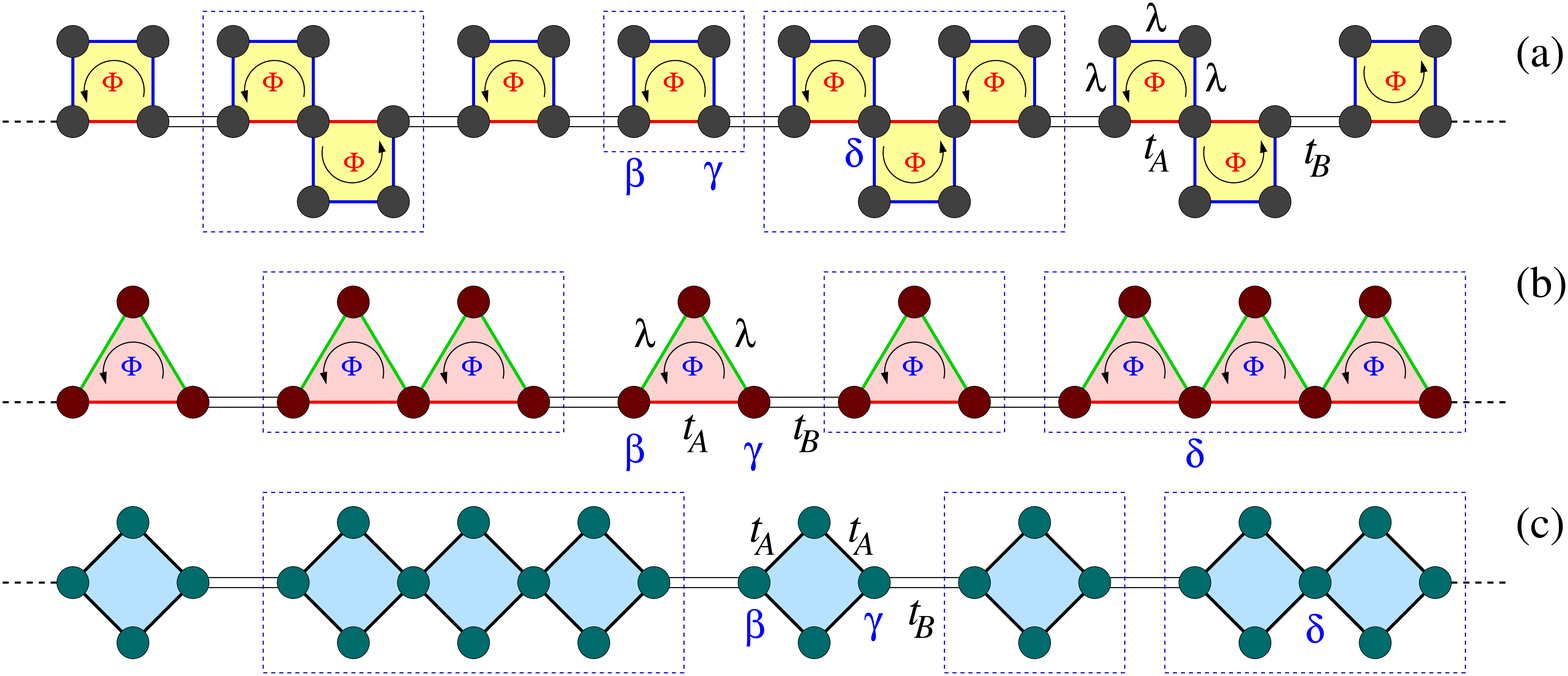}
\caption{(Color online) A typical random arrangement of the triangular and 
square plaquettes where two building blocks can even `touch' each other. There 
are now three kinds of sites along the backbone, viz., $\beta$, $\gamma$ and 
$\delta$, the last one having a coordination number four.}
\label{ringlattice}
\end{figure*}

Second important issue is the confirmation of the extended character of the eigenfunctions 
populating the continuous part of the LDOS spectrum in all the cases. At least, for the 
deterministic Fibonacci chain (or it's generalizations) an interesting answer to this 
question can be obtained by looking at how the on-site potentials and hopping integrals 
flow under successive RSRG iterations. Let's try to understand.
 
In the local coupling case for example, with reference to the Eq.~\eqref{rgrecur1}, it 
is found that, as soon as we set $\lambda = \sqrt{t_{B}^{2}-t_{A}^{2}}$ the parameter space 
follows the pattern $\epsilon_{\beta}(n)=\epsilon_{\gamma}(n)\ne \epsilon_{\alpha}(n)$ and 
$t_{A}(n) \ne t_{B}(n)$ at every $n$-th stage of renormalization, whenever we select an 
eigenvalue $E$ arbitrarily from within the two continuous subbands in the LDOS spectrum. 
This observation is substantiated by extensive numerical search throughout the observed 
continua scanned in arbitrarily small energy intervals. That such a pattern should 
correspond to extended Bloch-like eigenfunctions can be justified by considering 
Fig.~\ref{rgflow} 
where a perfectly periodic lattice of identical on-site potential $\epsilon$ and a 
constant nearest neighbor hopping $t$ is artificially converted into a golden mean 
Fibonacci chain. On this {\it artificial} Fibonacci chain 
$\epsilon_{\beta,0}=\epsilon_{\gamma,0}=\epsilon+t^{2}/(E-\epsilon)$, and different from  
$\epsilon_{\alpha,0}=\epsilon+2t^{2}/(E-\epsilon)$. At the same time, 
$t_{A,0}=2t^{2}/(E-\epsilon) \ne t_{B,0}$, the latter being equal to $t$. The 
{\it flow pattern} that we have been talking about therefore sets in at the very beginning.

The artificial Fibonacci chain in Fig.~\ref{rgflow} can now be renormalized using 
the recursion relations Eq.~\eqref{rgrecur1}, and the density of states may be obtained 
from the appropriate Green's function. As the parent lattice now is an ordered one, the 
typical one dimensional density of states is reproduced with the edges characterized by 
the van Hove singularity. The spectrum is absolutely continuous, and all the wave 
functions are Bloch functions. Interestingly, with the initial set of values as given 
above, the on site potentials and the hopping integrals for the scaled version of the 
{\it artificial} Fibonacci lattice (Fig.~\ref{rgflow}) get locked into the flow pattern 
$\epsilon_{\beta,n}=\epsilon_{\gamma,n} \ne \epsilon_{\alpha,n}$ and 
$t_{A,n} \ne t_{B,n}$ at every $n$-th stage of renormalization, and for all energy 
eigenvalues within the range $[\epsilon-2t,\epsilon+2t]$.

In our actual case of Fibonacci arrangement of the clusters in Fig.~\ref{cells}(a) 
as soon as such a flow pattern is set in for a special value of $\lambda$, it becomes 
impossible to judge whether the parent lattice was an ordered, perfectly periodic one, 
or a truly quasiperiodic Fibonacci chain. Thus the extendedness of the wave functions 
is firmly established whenever such an RSRG flow is observed. Same flow pattern is 
also observed in the cases of Fig.~\ref{cells}(b) and (c). In these cases, if we set 
beforehand $\lambda=t_{B}=t_{A}/\sqrt{2}$, then the {\it desired} flow of the 
parameters can be achieved by tuning the external flux to $\Phi=\Phi_{0}/4$. This refers 
to the interesting case of a {\it flux driven crossover} in the fundamental character 
of the wave functions in a non-locally coupled case or in the mixed case. In addition, 
for both the NLC case and the MC case, the LDOS at the $\beta$, $\gamma$ or $\alpha$ 
sites turn out to be exactly same whenever the resonance condition is satisfied. This 
is remarkable. The NLC and MC lattices are topologically different. An equality of the 
LDOS for $\lambda=t_{B}=t_{A}/\sqrt{2}$ and $\Phi=\Phi_{0}/4$ implies that for both 
these cases the parameters 
$(\epsilon_{\alpha},\epsilon_{\beta},\epsilon_{\gamma},t_{A},t_{B})$, initially 
represented by two different {\it points} (as their initial values are different) in 
the five dimensional parameter space, are driven to the same {\it fixed point} following 
two different trajectories.
 
As we have already mentioned, the result is independent of the order of arrangement 
of the triangles or the square boxes. Thus, for the same resonance condition an 
indefinite number of geometrically different systems, beginning their `journey' at different 
locations in the five dimensional parameter space finally flow, following different 
trajectories, to the same fixed point, and thus come under the common umbrella. We are 
tempted to conceptualize a kind of {\it universality class} from this point of view. 
It is to be noted however, that the comment is based on the observed LDOS at the 
sites on the backbone only. The average density of states can be different though. 
\section{Other geometries}
\label{othergeometries}
Before we end, it should be mentioned that, the central idea presented in the 
present work is not restricted to only the geometries discussed here. For 
example, one can have an array of triangular or square plaquettes without any 
isolated $\alpha$-site, where the plaquettes can `touch' each other giving rise 
to an additional site named $\delta$ and having a coordination number four.
We refer to Fig.~\ref{ringlattice} for a display of a disordered arrangement 
of such building blocks. The analysis proceeds in the same way and the one comes 
across a varied set of geometries for which the disorder-induced localization 
(or, a quasiperiodicity driven power law localization) can be suppressed and a full band 
(or subbands) of extended eigenfunctions can be generated.  
\section{Conclusion}
\label{conclu}
In conclusion, we have presented a class of topologically disordered array of 
building blocks described within a tight binding formalism, where delocalization 
of electronic eigenfunctions occur over either two subbands or over the entire range 
of allowed energies whenever the lattice parameters are inter-related through 
certain algebraic relation. We can have absolutely continuous spectrum even for 
such disordered or quasiperiodic arrangement of the unit cells in such cases. Even 
an external magnetic field can be used to delocalize the electronic states over a 
continuous band of energy eigenvalues in certain cases. This aspect leads to the 
possibility of a flux driven state transition in such low dimensional systems. 
\begin{acknowledgements}
Biplab Pal gratefully acknowledges a DST-INSPIRE Fellowship, and AC is thankful to 
DST, India for partial financial support through a PURSE programme of the 
University of Kalyani.
\end{acknowledgements}

\end{document}